\documentclass[12pt]{iopart}

\usepackage{graphicx}

\newcommand{\as}{\alpha_s}

\def\eq#1{{Eq.~(\ref{#1})}}
\def\fig#1{{Fig.~\ref{#1}}}

\def\peq#1{{(\ref{#1})}}

\begin{document}

\title{AdS/CFT applications to relativistic heavy ion collisions: a
  brief review}

\author{Yuri V. Kovchegov}

\address{Department of Physics, The Ohio State University, 191 West
  Woodruff Avenue, Columbus, OH 43210, USA} \ead{kovchegov.1@asc.ohio-state.edu}
\begin{abstract}
  We review some of the recent progress in our understanding of the
  physics of ultrarelativistic heavy ion collisions due to
  applications of AdS/CFT correspondence.
\end{abstract}



\section{Introduction}

Ultrarelativistic heavy ion collision experiments are being carried
out with the goal of creating a thermalized medium made out of
deconfined quarks and gluons, known as the quark--gluon plasma (QGP)
\cite{Harris:1996zx,Muller:2006ee,Muller:2012zq}.  The aim is to study
the properties of QGP, from those characterizing a static thermal
bath, such as the equation of state and the order of the phase
transition, to the dynamical properties, like the transport
coefficients, including shear and bulk viscosities of QCD matter.
With heavy ion experiments running at RHIC and LHC, qualitative and
quantitative theoretical progress is required both to better
understand the existing data and to suggest new observables to
measure.

Heavy ion collision is a multi-scale process, with the harder momenta
dominating the early times right after the collision, and the softer
momentum scales becoming relevant at the later time scales. Due to the
running of the QCD coupling constant the hard modes are
weakly-coupled, while the soft modes are coupled strongly. In recent
years a body of evidence has accumulated suggesting that the
non-perturbative large-coupling effects could be more important for
high-energy heavy ion collisions than was previously thought. Perhaps
the most important piece of evidence is the success of ideal
hydrodynamics in describing the evolution of the medium produced in
heavy ion collisions \cite{Kolb:2003dz,Teaney:2001av}. Corrections to
ideal hydrodynamics are inversely proportional to the coupling
constant and are small when the coupling is large. An example of this
is the shear viscosity, which (at weak coupling) scales as $\eta \sim
T^3/g^4 \ln (1/g)$ in a thermal QCD medium with the temperature $T$
\cite{Arnold:2000dr}: extrapolating this result one sees that the
shear viscosity is small when the coupling $g$ is large, generating a
small correction to the energy-momentum tensor of ideal hydrodynamics.
Another argument in favor of strong-coupling dynamics playing an
important role in heavy ion collisions is the large amount of jet
quenching needed to describe the data coming from RHIC and LHC. The
relevant jet quenching parameter $\hat q$, which is proportional to
the energy loss per unit path length of an energetic parton traversing
a medium \cite{Baier:1996sk}, was found to be rather large in some
analyses, possibly indicating that non-perturbative strongly-coupled
effects are at work \cite{Liu:2006ug}. Note however that $\hat q$ has
mass dimension of $M^3$, and the existence of a large momentum scale
${\hat q}^{1/3}$ likely indicates that at least a part of the process
is in fact weakly coupled, and, hence, perturbative. The last piece of
evidence suggesting possible importance of strong coupling effects in
the QCD plasma comes from lattice simulations, which indicate that the
energy density and pressure of the thermal QCD medium is about $80
\div 85 \%$ of that in an ideal gas of quarks and gluons for a broad
range of temperatures above the temperature of the QCD phase
transition $T_c$ \cite{Aoki:2005vt,Bazavov:2009zn}.  As the strong
coupling constant $\as (T)$ should get small at large-$T$, one expects
the high-temperature QCD medium to approach the Stefan-Boltzmann ideal
gas behavior as $T$ increases: the approach does take place, but is
much slower than expected by in the perturbative QCD calculations,
possibly indicating that non-perturbative effects are still important
even at reasonably high-$T$.  Note however that the resummed
perturbation series may successfully describe the lattice data
\cite{Andersen:2011sf}.

While none of the arguments presented above definitively demonstrates
dominance of strong-coupling effects in heavy ion collisions at RHIC
and LHC, we will proceed under the assumption that the strong-coupling
effects are important and try to analyze heavy ion collisions assuming
that the coupling is large. One reason for this is that exploring
heavy ion collisions in the strong-coupling regime is very interesting
by itself, providing a new theoretical angle on the process. In
addition, it may be that the typical strong coupling constant in these
collisions is neither very small nor very large, such that both small-
and large-coupling approaches would have some chance of describing the
data.

The strong-coupling QCD analysis of such an ultrarelativistic process
as a heavy ion collision is impossible with the present state of the
QCD theory. Instead an opportunity to study the strongly-coupled field
theories came up due to advances in string theory, culminating with
the formulation of the Anti-de Sitter space/Conformal Field Theory
(AdS/CFT) correspondence
\cite{Maldacena:1997re,Gubser:1998bc,Witten:1998qj,Witten:1998zw,Aharony:1999ti},
which is the duality between the ${\cal N} =4$ $SU(N_c)$ super
Yang-Mills (SYM) theory in four flat space-time dimensions (conformal
field theory, or simply CFT) and the type IIB superstring theory on
AdS$_5 \times$S$^5$. In the limit of large number of colors $N_c$ and
large 't Hooft coupling $\lambda = g^2 N_c$ such that $N_c \gg \lambda
\gg 1$, AdS/CFT correspondence reduces to the gauge-gravity duality:
${\cal N} =4$ SU($N_c$) SYM theory at $N_c \gg \lambda \gg 1$ is dual
to the (weakly coupled) classical supergravity in AdS$_5 \times$S$^5$.
The AdS/CFT correspondence gives us a powerful tool one can use to
systematically study a strongly-coupled gauge theory (specifically,
${\cal N} =4$ SYM theory).  Using the AdS/CFT correspondence one can try
to study heavy ion collisions at strong coupling. Indeed one should be
mindful about the many differences between QCD and the ${\cal N} =4$
SYM theory: the former is confining, while the latter is not; QCD
coupling runs, while it is a constant in the conformal ${\cal N} =4$
SYM theory; QCD plasma has a chiral phase transition, while ${\cal N}
=4$ SYM plasma does not.  Until the errors introduced by using a
theory different from QCD are understood, any conclusions we derive
from applying the AdS/CFT correspondence to the study of QCD-mediated
processes should be taken as qualitative at best. However, in some
cases one may hope that AdS/CFT predictions are universally valid for
all strongly-coupled theories, including QCD. This hypothesis is
supported by some numerical successes of AdS/CFT, an example of one of
which will be shown below.


\section{Elements of AdS/CFT correspondence and the strongly-coupled supersymmetric plasma}

For reviews of the AdS/CFT correspondence we refer the reader to
\cite{Aharony:1999ti,Klebanov:2000me,Kiritsis:2007zz,Son:2007vk}.
Since our goals in this mini-review are rather applied, we will simply
state the tools we need to accomplish them, without presenting any
proofs, and illustrate these statements by analyzing some of the
properties of the supersymmetric plasma.

AdS$_5 \times$S$^5$ space is obtained by stacking $N_c$ parallel D3
branes on top of each other. The resulting metric for (the Poincare
wedge of) AdS$_5 \times$S$^5$ in the $z \gg L$ limit is
\begin{equation}\label{AdSmetric}
  ds^2 = \frac{L^2}{z^2} \, \left[ - dt^2 + d {\vec x}^{\, 2} + d z^2
  \right] + L^2 \, d \Omega_5^2
\end{equation}
where $(t, {\vec x}) = (t, x^1, x^2, x^3)$ are the regular
four-dimensional coordinates, $z \in [0, +\infty)$ is the fifth
dimension of AdS$_5$, and the branes are located at $z = \infty$. $L$
is the radius of the S$^5$ sphere. For our purposes it is important to
note that the AdS$_5$ metric in \eq{AdSmetric} (everything on the
right except for the last term) satisfies Einstein equations
\begin{equation}
  \label{Einstein}
  R_{MN} - \frac{1}{2} \, g_{MN} \, R + \Lambda \, g_{MN} =0
\end{equation}
with $M, N = 0, \ldots , 4$ and the cosmological constant $\Lambda = -
6/L^2$.

The AdS/CFT correspondence is the statement of equivalence of the
${\cal N} =4$ SU($N_c$) SYM theory in four space-time dimensions and
the type IIB superstring theory on AdS$_5 \times$S$^5$
\cite{Maldacena:1997re}. For practical calculations the duality means
the following. Each operator $\cal O$ in the four-dimensional CFT is
paired up with the dual field $\varphi$ in the AdS$_5$ bulk. Then the
duality can be mathematically formulated as
\cite{Gubser:1998bc,Witten:1998qj,Witten:1998zw}:
\begin{equation}
  \label{duality}
  \left\langle \exp \left[ \int d^4 x \, {\cal O} (x) \, \varphi_0 (x) \right] 
  \right\rangle_{\mbox{CFT}_4} = Z_{\mbox{string}} \left[ 
   \varphi (x, z) \bigg|_{z \approx 0} = z^{4 - \Delta} \, \varphi_0 (x) \right].
\end{equation}
On the left hand side of \eq{duality} we have an expectation value
taken in the full four-dimensional CFT, while on the right there is a
partition function for the type IIB superstring theory, calculated
with a particular boundary condition on the bulk field $\varphi (x,
z)$.  $\Delta$ is the scaling dimension of the operator ${\cal O}$.
One can show that in the $N_c \gg \lambda \gg 1$ limit the string
partition function becomes classical, $Z_{\mbox{string}} \approx e^{i
  \, S_{bulk}}$ with $S_{bulk}$ the classical action on AdS$_5
\times$S$^5$ calculated with the $\varphi (x, z) \big|_{z \approx 0} =
z^{4 - \Delta} \, \varphi_0 (x)$ boundary condition. Thus the
incredibly complicated quantum field theoretical problem of
calculating an expectation value in the full CFT is reduced to a
classical calculation on AdS$_5 \times$S$^5$. This is a great
simplification due to the AdS/CFT correspondence.

\begin{figure}[ht]
\begin{center}
\includegraphics[width=0.6\textwidth]{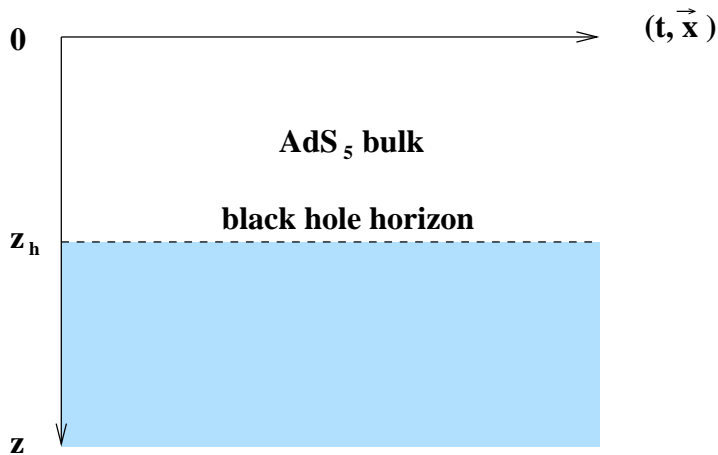}
\end{center}
\caption{Pictorial illustration of the AdS Schwarzschild black hole metric \peq{AdSSBH}.}
\label{AdSSBH_fig}
\end{figure}

As an important example of the use of AdS/CFT duality which is also
useful for heavy ion collisions, let us consider a thermal medium made
out of strongly-coupled ${\cal N} =4$ SYM matter. Due to thermodynamic
properties of black holes, the dual of the thermal medium in the gauge
theory is the AdS Schwarzschild black hole (AdSSBH) with the metric
\cite{Maldacena:1997re}
\begin{equation}\label{AdSSBH}
  ds^2 = \frac{L^2}{z^2}\left[ - \left( 1 - \frac{z^4}{z^4_h} \right) \,
    d t^2 + d {\vec x}^{\, 2} + \frac{dz^2}{1 - \frac{z^4}{z^4_h}}  \right].
\end{equation}
The horizon of AdSSBH metric is located at $z = z_h$. Introducing the
Euclidean time $t_E = i \, t$ the metric \peq{AdSSBH} can be extended
to the Euclidean space-time.  Expand the Euclidean version of the
metric \peq{AdSSBH} near the horizon by redefining $z = z_h (1 -
\rho^2/L^2)$: at the lowest non-trivial order in $\rho$ we get
\begin{equation}
  \label{near_h}
  ds^2 \approx \frac{4 \, \rho^2}{z_h^2} \, d t_E^2 + \frac{L^2}{z_h^2} \, 
 d {\vec x}^{\, 2} + d \rho^2.
\end{equation}
We see that the $(t_E, \rho)$-part of the metric looks like the polar
coordinates in 2d, with the metric $d s^2_{2d} = \rho^2 \, d \phi^2 +
d \rho^2$, if we define the ``angle'' $\phi = 2 \, t_E/z_h$. The
metric $d s^2_{2d}$ has no kink singularity at $\rho =0$ only if it is
periodic in $\phi$ with the period of $2 \, \pi$. This means that the
Euclidean time $t_E$ should be periodic with the period of $\Delta t_E
= \pi \, z_h$. Remembering that a thermal field theory in Euclidean
space is periodic with the period $\Delta t_E = \beta = 1/T$ with $T$
the temperature, we (i) find the Hawking temperature $T_H = 1/(\pi
z_h)$ of the AdS Schwarzschild black hole \peq{AdSSBH} and (ii)
identifying the Hawking temperature of AdSSBH with the temperature of
the dual gauge theory we obtain
\begin{equation}
  \label{zhT}
  z_h = \frac{1}{\pi \, T}. 
\end{equation}
The metric of \peq{AdSSBH} is illustrated in \fig{AdSSBH_fig}. We see
from \eq{zhT} that at high temperature $T$ the horizon is close to the
boundary of the AdS space at $z=0$, while for low $T$ the horizon is
deep in the AdS$_5$ bulk.

\begin{figure}[ht]
\begin{center}
\includegraphics[width=0.55\textwidth]{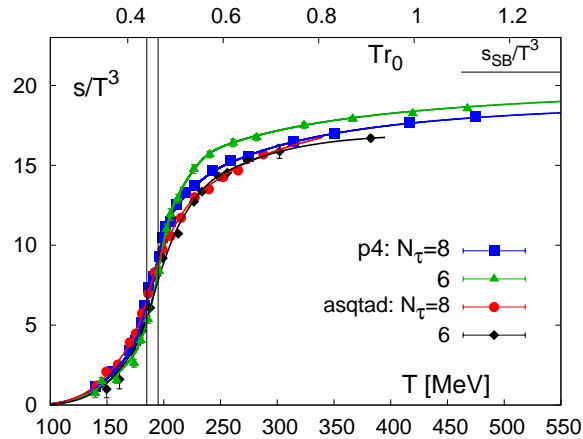}
\end{center}
\caption{Entropy density $s$ (scaled down by $T^3$) of the (2+1)-flavor 
  QCD as a function of temperature determined by the lattice
  simulations in \cite{Bazavov:2009zn}. The ideal gas Stefan-Boltzmann
  limit is shown by a horizontal line. (Reprinted figure with
  permission from \cite{Bazavov:2009zn}. Copyright 2009 by the
  American Physical Society.)}
\label{entropySB}
\end{figure}

Identifying the AdS Schwarzschild black hole as the gravity dual of
the strongly-couples SYM plasma allows us to calculate the entropy of
the SYM plasma by equating it with the Bekenstein-Hawking entropy of a
black hole
\begin{equation}
  \label{BH}
  S_{BH} = \frac{A}{4 \, G_N}
\end{equation}
where $A$ is the horizon area and $G_N$ is the Newton's constant.
Since the area of a 5-sphere is $A_{S^5} = \pi^3 \, L^5$, the horizon
area for AdSSBH can be calculated via
\begin{equation}
  \label{hor_area}
  A = \int d x^1 \, d x^2 \, d x^3 \, \sqrt{\mbox{det} \, g_{3d} \big|_{z=z_h}} \ A_{S^5}
\end{equation}
where $\mbox{det} \,g_{3d} = L^6/z^6$ is the determinant of the part
of the metric \peq{AdSSBH} in three dimensions $(x^1, x^2, x^3)$.
Using \eq{zhT} we get $A = L^8 \, V_{3d} \, \pi^6 \, T^3$ with
$V_{3d}$ the 3d volume occupied by the plasma. Using this area in
\eq{BH} along with
\begin{equation}
  \label{GN}
  G_N = \frac{\pi^4 \, L^8}{2 \, N_c^2}
\end{equation}
we obtain the Bekenstein-Hawking entropy of AdSSBH, or, by AdS/CFT
duality, the entropy of the strongly-coupled ${\cal N} =4$ SYM plasma
\cite{Gubser:1996de}
\begin{equation}
  \label{entr_strong}
  S = \frac{\pi^2}{2} \, N_c^2 \, T^3 \, V_{3d}. 
\end{equation}
We have derived an expression for the entropy of a strongly-coupled
medium in a thermal field theory by doing a very simple gravity
calculation! Indeed the strong-coupling expression \peq{entr_strong}
corresponds to a sum of an infinite number of Feynman diagrams:
AdS/CFT duality allowed us to arrive at the answer in a much simpler
way.

\eq{entr_strong} should be compared to the Stefan-Boltzmann entropy of
an ideal gas of non-interacting particles in ${\cal N} =4$ SYM theory
\begin{equation}
  \label{entr_ideal}
  S_{SB} = \frac{2 \, \pi^2}{3} \, N_c^2 \, T^3 \, V_{3d}.
\end{equation}
We see that $S/S_{SB} = 3/4$: the entropy of a strongly-coupled
supersymmetric plasma is $3/4$ of that for the ideal gas. As was
mentioned before, this result seems to be in a semi-quantitative
agreement with the lattice QCD calculations for entropy above $T_c$
giving $S/S_{SB} = 80 \div 85 \%$ \cite{Aoki:2005vt,Bazavov:2009zn},
as illustrated in \fig{entropySB} for the entropy density $s =
S/V_{3d}$. Note, again, that in interpreting this semi-agreement one
has to remember that ${\cal N} =4$ SYM and QCD are different theories,
and the ${\cal N} =4$ SYM prediction \peq{entr_strong} for $S/T^3$
would correspond to a horizontal straight line in the plot of
\fig{entropySB} exhibiting no temperature dependence.

Now suppose we want to find the expectation value of the
energy-momentum tensor (EMT) $\langle T_{\mu\nu} \rangle$ in the
${\cal N} =4$ SYM plasma ($\mu,\nu = 0, \ldots , 3$ label the 4d
space-time). Since gravity couples to the EMT of the matter fields in
4d, by AdS/CFT rules \peq{duality} the EMT operator is dual to the
metric tensor $g_{MN}$ in the bulk \cite{deHaro:2000xn}. The AdS/CFT
duality \peq{duality} leads to the following prescription for the
calculation of $\langle T_{\mu\nu} \rangle$, known as the holographic
renormalization \cite{deHaro:2000xn}. First cast the metric in the
Fefferman-Graham form \cite{F-G}
\begin{equation}
  \label{FGmetric}
  d s^2 = \frac{L^2}{z^2} \, \left[ g_{\mu\nu} d x^\mu d x^\nu + d z^2 \right].
\end{equation}
Then expand the 4d metric $g_{\mu\nu} (x, z)$ near the boundary of
AdS$_5$ at $z=0$. One can show that if the expansion starts with the
Minkowski metric $\eta_{\mu\nu}$ at $z=0$ (as is the case for AdS$_5$)
then the next term in the expansion is order-$z^4$, such that
\cite{deHaro:2000xn}
\begin{equation}
  \label{mexp}
  g_{\mu\nu} (x, z) = \eta_{\mu\nu} + z^4 \, g_{\mu\nu}^{(4)} (x) + \ldots . 
\end{equation}
With the expansion \peq{mexp} in hand one can find the expectation
value of the EMT by the following simple relation
\begin{equation}
  \label{EMT1}
  \langle T_{\mu\nu} \rangle = \frac{N_c^2}{2 \, \pi^2} \ g_{\mu\nu}^{(4)} (x). 
\end{equation}

Applying this prescription to the AdSSBH metric \peq{AdSSBH} we first
perform a substitution
\begin{equation}
  \label{sub}
  z = \frac{{\tilde z}}{\sqrt{1 + \frac{{\tilde z}^4}{{\tilde z}_h^4}}}
\end{equation}
with ${\tilde z}_h = z_h \, \sqrt{2}$ recasting \eq{AdSSBH} into
Fefferman-Graham form \cite{Janik:2005zt}
\begin{equation}\label{AdSSBH_FG}
  ds^2 = \frac{L^2}{{\tilde z}^2}\left[ - \frac{\left( 1 - \frac{{\tilde z}^4}{{\tilde z}^4_h} 
\right)^2}{1 + \frac{{\tilde z}^4}{{\tilde z}^4_h}} \,
    d t^2 + \left( 1 + \frac{{\tilde z}^4}{{\tilde z}^4_h} \right) \, 
d {\vec x}^{\, 2} + d {\tilde z}^2  \right].
\end{equation}
Reading off the $z^4$ coefficients of $g_{\mu\nu} (x, z)$ from
\eq{AdSSBH_FG} we obtain the EMT of a static supersymmetric plasma
\begin{equation}
  \label{EMT_plasma}
  \langle T_{\mu\nu} \rangle = \frac{\pi^2}{8} \, N_c^2 \, T^4 \, \mbox{diag} 
  \left\{ 3 , 1, 1, 1 \right\}. 
\end{equation}
As we will see below, holographic renormalization is a powerful tool
that can be used to study heavy ion collisions at strong coupling.


\section{Heavy ion collisions in AdS$_5$}

As we mentioned above, heavy ion collision is a multi-scale process
involving an interplay of small- and large-coupling phenomena.
Unfortunately no single theoretical framework incorporates both
perturbative and non-perturbative effects. Therefore, applying AdS/CFT
duality to heavy ion collisions one has to make a choice of either
treating the whole collision as a strongly-coupled process, or,
perhaps more realistically, limit the application of AdS/CFT duality
to the description of the produced medium at the later times after the
collision when the QCD coupling is more likely to be large. In this
Section we will discuss the progress in the former direction, while
the advances in the latter one are presented in the next Section.


\subsection{Shock wave collisions}

\begin{figure}[h]
  \begin{center}
    \includegraphics[width=6cm]{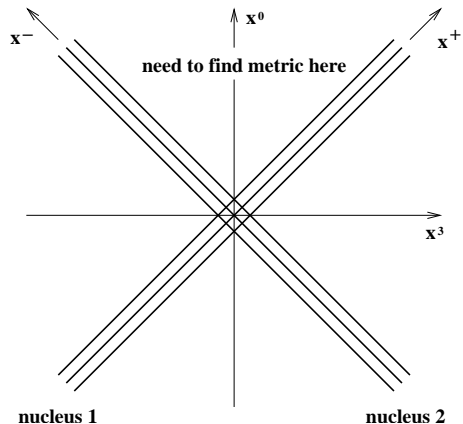}
  \end{center}
  \caption{The space-time picture of the ultrarelativistic heavy ion 
    collision in the center-of-mass frame. The collision axis is
    labeled $x^3$, the time is $x^0$.}
  \label{spacetime}
\end{figure}

We begin by modeling the whole heavy ion collision in the
strongly-coupled framework of the AdS/CFT correspondence. The goal
here is rather ambitious, and includes understanding particle
production in the collision, followed by the thermalization of the
produced matter, along with the subsequent evolution of the resulting
thermal medium: essentially one wants to understand the whole
collision in a single unified framework. While there is no nuclei (or
other bound states) that we could imagine colliding in the conformal
${\cal N} =4$ SYM theory, one can follow Bjorken \cite{Bjorken:1982qr}
and model the colliding nuclei as the Lorentz-contracted thin sheets
of matter, which are uniform and extend to infinity in the transverse
direction.  The collision of two such ``nuclei'' is illustrated in
\fig{spacetime} using light-cone coordinates $x^\pm = (x^0 \pm
x^3)/\sqrt{2}$.

The nucleus moving along the $x^+$-axis (``nucleus 1'' in
\fig{spacetime}) has a large $T^{++}$ component of its EMT, with all
other EMT components negligibly small. Moreover, due to assumptions we
made about the uniform distribution in the transverse plane, $T^{++}$
is a function of $x^-$ only with its support localized near $x^- =0$
due to Lorentz contraction. The metric dual to such an
ultrarelativistic nucleus which solved Einstein equations
\peq{Einstein} is that of a gravity shock wave in the bulk
\cite{Janik:2005zt}
\begin{equation}\label{1nuc}
  ds^2 \, = \, \frac{L^2}{z^2} \, \left\{ -2 \, dx^+ \, dx^- + \frac{2
      \, \pi^2}{N_c^2} \, \langle T^{++} (x^-) \rangle \, z^4 \, d
    x^{- \, 2} + d x_\perp^2 + d z^2 \right\}
\end{equation}
with $d x_\perp^2 = (d x^1)^2 + (d x^2)^2$.  The metric dual to
``nucleus 2'' in \fig{spacetime} is obtained from \eq{1nuc} by the $+
\leftrightarrow -$ interchange. 

The problem of understanding heavy ion collisions in this version of
AdS/CFT approach can be formulated as follows: we know the metric for
$x^+ < 0$ or for $x^- < 0$, where it is simply a sum of the two shock
wave metrics [\eq{1nuc} and the one obtained from it by $+
\leftrightarrow -$]. What is the metric solving Einstein equations
\peq{Einstein} at $x^+, x^- \ge 0$, i.e., in the forward light-cone
region, as shown in \fig{spacetime}?

To date there exists no analytical solution of this problem. On the
analytical side the problem was tackled perturbatively order-by-order in
the EMT's of both shock waves \cite{Grumiller:2008va,Albacete:2008vs}
and by an all-order resummation of the perturbation theory in the EMT
of one of the shocks \cite{Albacete:2009ji}. One usually approximates
the nuclear profiles by delta-functions: defining $t_1 (x^-) \, \equiv
\, \frac{2 \, \pi^2}{N_c^2} \, \langle T_{1}^{++} (x^-) \rangle$ and
$t_2 (x^+) \, \equiv \, \frac{2 \, \pi^2}{N_c^2} \, \langle T_{2}^{--}
(x^+) \rangle$ for the EMT's of the two shock waves, we have
\begin{equation}\label{deltas}
  t_1 (x^-) = \mu_1 \, \delta (x^-), \ \ \ t_2 (x^+) = \mu_2 \, \delta
  (x^+)
\end{equation}
with \cite{Albacete:2008vs,Albacete:2008ze}
\begin{equation}\label{mus}
  \mu_{1} \sim p_{1}^+ \, \Lambda_1^2 \, A_1^{1/3}, \ \ \ \mu_{2} \sim
  p_{2}^- \, \Lambda_2^2 \, A_2^{1/3}.
\end{equation}
Here $p_1^+$, $p_2^-$ are the large light-cone momenta per nucleon in
the two nuclei with atomic numbers $A_1$ and $A_2$, while $\Lambda_1$
and $\Lambda_2$ are the typical transverse momentum scales describing
the two nuclei. 

The metric in the forward light-cone can be represented as a series in
the powers of $\mu_1$ and $\mu_2$, corresponding to dimensionless
Lorentz-invariant expansion parameters $\mu_1 \, ( x^- )^2 \, x^+$ and
$\mu_2 \, ( x^+ )^2 \, x^-$. The idea of solving Einstein equations by
an expansion in powers of $\mu_1$ and $\mu_2$ was formulated in
\cite{Grumiller:2008va} for the delta-function profiles \peq{deltas}
and extended to other more realistic longitudinal profiles in
\cite{Albacete:2008vs} and to non-trivial transverse profiles in
\cite{Taliotis:2010pi}.  The all-order resummation of the powers of
$\mu_2$ keeping $\mu_1$ at the lowest non-trivial order was carried
out in \cite{Albacete:2009ji} resulting in the following expectation
value of the EMT in the forward light-cone (only non-zero components
are listed): \numparts
\begin{eqnarray}
  \langle T^{++}\rangle \, = \, - \frac{N_c^2}{2 \, \pi^2} \,
  \frac{4 \, \mu_1 \, \mu_2 \, (x^+)^2 \, \theta (x^+) \, \theta
    (x^-)}{\left[ 1 + 8 \, \mu_2 \, (x^+)^2 \, x^- \right]^{3/2}}, \label{T++} \\
  \langle T^{--}\rangle \, = \, \frac{N_c^2}{2 \, \pi^2} \, \theta
  (x^+) \, \theta (x^-) \, \frac{\mu_1}{2 \, \mu_2 \, (x^+)^4} \,
  \frac{1}{\left[ 1 + 8 \, \mu_2 \, (x^+)^2 \, x^- \right]^{3/2}}
   \nonumber \\ \times \, \bigg[ 3 - 3 \, \sqrt{1 + 8 \, \mu_2 \,
    (x^+)^2 \, x^-}  + 4 \, \mu_2 \, (x^+)^2 \, x^- \nonumber \\ \times
  \, \left( 9 + 16 \, \mu_2 \, (x^+)^2 \, x^- - 6 \, \sqrt{1 + 8 \,
      \mu_2 \, (x^+)^2 \, x^-}
  \right) \bigg] , \label{T--} \\
  \langle T^{+-}\rangle \, = \, \langle T^{11}\rangle \, = \, 
\langle T^{22}\rangle \, = \, \frac{N_c^2}{2 \, \pi^2} \, \frac{8
    \, \mu_1 \, \mu_2 \, x^+ \, x^- \, \theta (x^+) \, \theta
    (x^-)}{\left[ 1 + 8 \, \mu_2 \, (x^+)^2 \, x^- \right]^{3/2}}, \label{T+-}
\end{eqnarray}
\endnumparts which, by construction, is applicable in the region
defined by $\mu_1 \, ( x^- )^2 \, x^+ \ll 1$. Due to this limitation,
the EMT in Eqs.~\peq{T++}, \peq{T--}, and \peq{T+-} is not valid at
asymptotically late proper times $\tau = \sqrt{2 \, x^+ \, x^-}$ when
the system is likely to thermalize and should be described by ideal
hydrodynamics: it does not yield us a new solution of ideal
hydrodynamics equations. (For instance, $\langle T^{++}\rangle < 0$ in
\eq{T++}, which is impossible in ideal hydrodynamics.) Rather
Eqs.~\peq{T++}, \peq{T--}, and \peq{T+-} describe a non-equilibrium
medium on its way to equilibration.

\begin{figure}[t]
  \begin{center}
    \includegraphics[width=8cm]{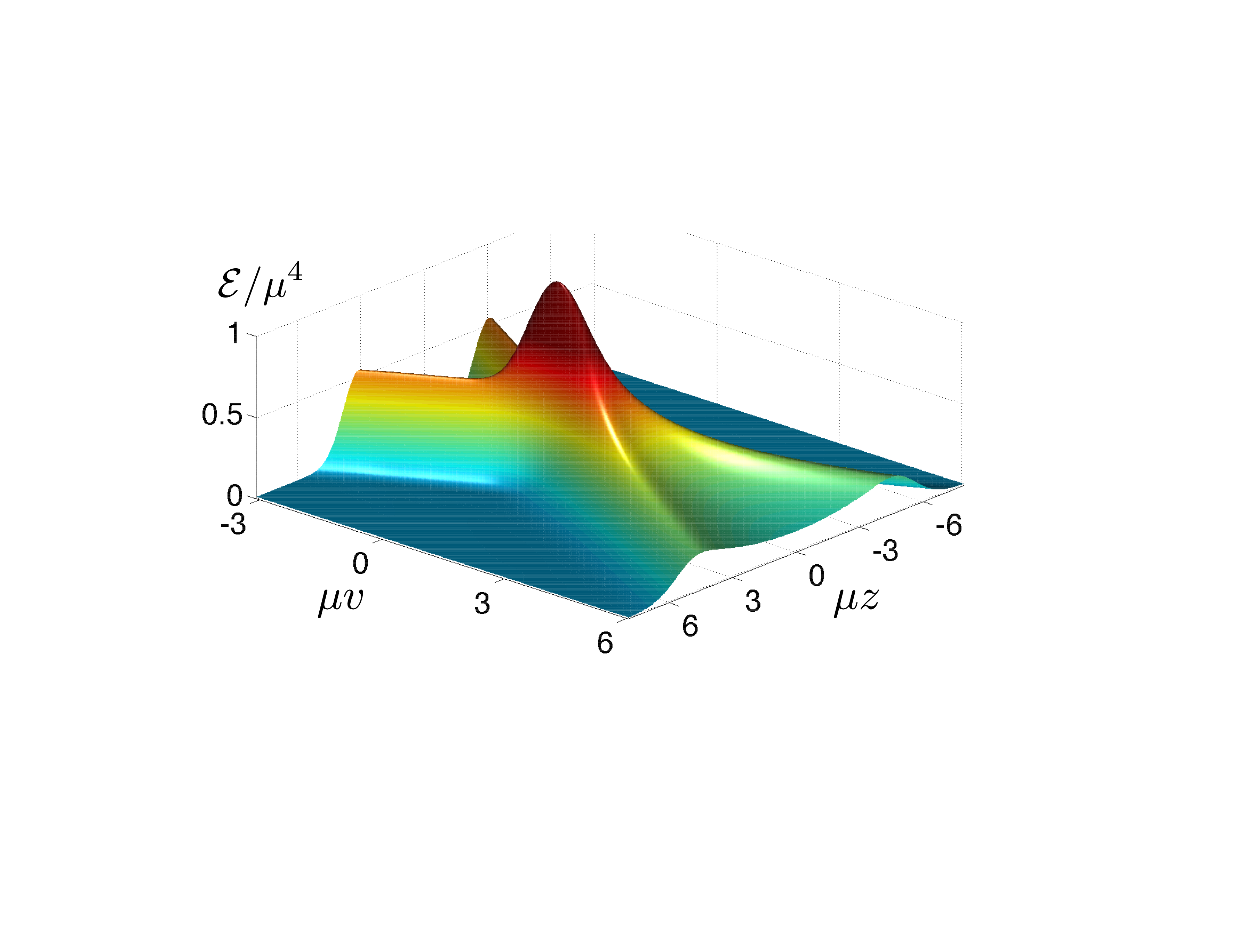}
  \end{center}
  \caption{Energy density in the shock wave collision obtained by the 
    numerical solution of the problem carried out in
    \cite{Chesler:2010bi}. In the Eddington-Finkelstein coordinates
    used in \cite{Chesler:2010bi}, $v$ is time, while $z$ is the
    longitudinal coordinate (the collision axis). $\mu$ is a parameter
    with dimension of a mass, which is related but is not equal to
    $\mu_1, \mu_2$ introduced in the text. (Reprinted figure with
    permission from \cite{Chesler:2010bi}. Copyright 2011 by the
    American Physical Society.)}
  \label{ED}
\end{figure}

The all-order resummation of $\mu_2 \, ( x^+ )^2 \, x^-$ in
\cite{Albacete:2009ji} allowed for one more insight about shock wave
collisions: smearing the profile of the first shock wave to generate a
finite width $a_1$
\begin{equation}
  \label{t1s}
  t_1 (x^-) \, = \, \frac{\mu_1}{a_1} \, \theta (x^-) \,
  \theta (a_1 - x^-)
\end{equation}
one can follow the evolution of the shock wave after the collision,
which yields \cite{Albacete:2009ji}
\begin{equation}\label{stop_pA}
  \langle T^{++} \rangle \, = \, \frac{N_c^2}{2 \, \pi^2} \,
  \frac{\mu_1}{a_1} \, \frac{1}{\sqrt{1 + 8 \, \mu_2 \, (x^+)^2 \,
      x^-}}, \ \ \ \mbox{for} \ \ \ 0 < x^- < a_1, \ \ x^+ > 0.
\end{equation}
We see that $\langle T^{++} \rangle \rightarrow 0$ with increasing
$x^+$. This implies that on the time scales of the order of
$x^+_{stop} \, \sim \, \frac{1}{\sqrt{\mu_2 \, a_1}}$ the shock wave
would lose much of its light cone momentum and would ``stop'', i.e.,
it would either deviate from its light-cone trajectory or would simply
dissolve into the produced medium. This ``stopping'' effect may be due
to the following mechanism: the partons in the colliding nuclei branch
unchecked due to the large value of the coupling, which results in the
shock waves consisting of a huge number of very low-momentum partons.
When two systems of low-momentum partons collide, the partons easily
slow each other down, resulting in the ``stopping'' of \eq{stop_pA}.
This ``stopping'' conclusion, unfortunately, is a strong argument
against using AdS/CFT to describe the whole heavy ion collision, since
no such ``stopping'' behavior appears to have been observed
experimentally at RHIC and LHC.

A numerical solution of the shock wave collision problem was carried
out in \cite{Chesler:2010bi} for the shock waves with Gaussian
profiles (see also \cite{Wu:2011yd}). The energy density obtained in
\cite{Chesler:2010bi} is plotted in \fig{ED}. The results of
\cite{Chesler:2010bi} appear (at least qualitatively) to confirm the
above analytical conclusion about the fast depletion of the EMT along
the shock wave's light-cone \peq{stop_pA}.  In addition, the numerical
simulation suggests a rather fast onset of viscous hydrodynamics,
indicating a fast equilibration of the produced medium.


\subsection{Trapped surface analysis and thermalization}

Interestingly enough, one does not have to solve Einstein equations
explicitly to figure our the fate of the colliding shock waves system.
It is possible to determine whether the system reaches thermal
equilibrium, that is whether a black hole is created in the bulk, by
performing a trapped surface analysis \cite{Penrose,Eardley:2002re}.
According to the Hawking-Penrose theorem, existence of a trapped
surface implies that a gravitational collapse (and, hence, a black
hole creation) is inevitable. A trapped surface analysis for colliding
shock waves in AdS$_5$ was carried out in
\cite{Gubser:2008pc,Lin:2009pn,Gubser:2009sx,Kovchegov:2009du}, with
the original work on the subject \cite{Gubser:2008pc} dealing with
shock waves with non-trivial transverse coordinate profiles, obtained
by putting ultrarelativistic point sources in the bulk. We now know
that trapped surfaces are formed in collisions of a variety of shock
waves with and without sources in the bulk
\cite{Gubser:2008pc,Lin:2009pn,Gubser:2009sx,Kovchegov:2009du},
indicating that thermalization does take place in heavy ion collisions
at strong coupling.

The area of the trapped surface gives one the lower bound on the
entropy of the black hole, and hence, on the entropy of the produced
matter. For the sourceless delta-function shock waves of Eqs.
\peq{1nuc}, \peq{deltas} the produced entropy per unit transverse area
$A_\perp$ was found to be \cite{Kovchegov:2009du}
\begin{equation}
  \frac{S}{A_\perp} \, = \, \frac{N_c^2}{2 \, \pi^2} \, \left( 2 \,
    \mu_1 \, \mu_2 \right)^{1/3}.
\end{equation}
The thermalization (black hole formation) proper time is given by the
only dimensionful boost-invariant parameter in the problem ($\mu_1 \,
\mu_2$) \cite{Grumiller:2008va,Kovchegov:2009du} as
\begin{equation}
  \label{thermtime}
  \tau_{th} \, \sim \, \frac{1}{(\mu_1 \, \mu_2)^{1/6}},
\end{equation}
which, as can be seen from \eq{mus}, is very short parametrically,
being suppressed by a power of the center of mass energy $s$ as $\mu_1
\, \mu_2 \sim p_1^+ \, p_2^- \sim s$. This conclusion was supported by
the numerical simulations of \cite{Chesler:2010bi} performed after
\cite{Grumiller:2008va,Kovchegov:2009du}.  Unfortunately, substituting
realistic values for $\mu_1, \mu_2$ into \eq{thermtime} leads to the
thermalization time which is too short to agree with the hydrodynamic
analyses of the RHIC and LHC data
\cite{Grumiller:2008va,Kovchegov:2009du}.

It is worth noting here that the numerical simulations of
\cite{Chesler:2010bi} give $0.35$~fm/c as the time interval between
when the Gaussian shocks used in the simulations start to overlap
significantly to the onset of viscous hydrodynamics in RHIC
kinematics. This number is somewhat closer to the time scale on the
order of $1 - 2$~fm/c used in hydrodynamic simulations
\cite{Teaney:2003kp,Romatschke:2007mq,Song:2010mg,Qiu:2011hf} than the
time scale of about $\tau_{th} \approx 0.1$~fm/c resulting from
\eq{thermtime} evaluated for RHIC kinematics. The difference between
the two numbers may be due in part to the finite longitudinal width of
colliding shock waves used in the numerical simulations
\cite{Chesler:2010bi} versus the zero effective width of the
delta-function shock waves \peq{deltas}. However, note that to
properly compare thermalization time with the hydrodynamic simulations
\cite{Teaney:2003kp,Romatschke:2007mq,Song:2010mg,Qiu:2011hf} one has
to start measuring the proper time from $\tau =0$, i.e., from the
moment of the complete overlap of the two shocks: such procedure, when
applied to the results of \cite{Chesler:2010bi}, yields the
thermalization time of about $\tau_{th} \approx 0.17$~fm/c, much
closer to \eq{thermtime} and further away from the phenomenological
times.

Finally, noting again that $\mu_1 \, \mu_2 \sim s$ with $s$ the center
of mass energy of the collision, we get \cite{Gubser:2008pc}
\begin{equation}
  \frac{S}{A_\perp} \, \propto \, s^{1/3}. 
\end{equation}
Identifying the produced entropy with the number of degrees of
freedom, and hence with the number of hadrons produced in the
collision, we conclude that the AdS/CFT prediction is that the hadron
multiplicity should grow as $dN/d\eta \sim s^{1/3}$ with $\eta$ the
pseudo-rapidity. Unfortunately the experimental data gives a much
smaller power of energy, $dN/d\eta \sim s^{0.15}$
\cite{Aamodt:2010pb}, possibly signaling again that the early stages
of heavy ion collisions (in which the majority of the entropy is
produced) are not strongly-coupled. Note, however, that modifications
of AdS$_5$ geometry allow for a better agreement with the data
\cite{Kiritsis:2011yn}.


\section{Evolution of the produced medium: hydrodynamics and the shear viscosity bound}

Let us now change the strategy and assume that the actual collision is
either weakly-coupled or is described by the physics outside the
gauge-gravity duality for some other reasons. Strong-coupling physics
in this scenario becomes dominant at some later (though not
necessarily very late) proper time $\tau = \sqrt{2 \, x^+ \, x^-}$ in
the collision. We are thus confined to the forward light-cone of
\fig{spacetime} without specific information about the past of the
system (that is, without the shock waves). The question now is whether
we can say anything specific about the strong-coupling dynamics of
such a medium.

Assuming again that the system is uniform in the transverse direction,
we see that the dynamics in the forward light cone in 4d depends only
on two variables: the proper time and the space-time rapidity
$\eta_{st} = (1/2) \ln (x_+/x_-)$. Furthermore, let us assume for
simplicity that the matter distribution is $\eta_{st}$-independent
\cite{Bjorken:1982qr}: the assumption is justified by the weak
rapidity-dependence of the particle multiplicity near mid-rapidity in
the actual heavy ion experiments. The most general metric in AdS$_5$
describing $\eta_{st}$-independent medium in the forward light cone in
the Feffermann-Graham coordinates is \cite{Janik:2005zt}\footnote{Note
  that the bulk $z$-variable in Feffermann-Graham coordinates is, in
  general, different from $z$ used in, say, \eq{AdSSBH}: for that
  particular black hole metric the relation is given by \eq{sub}.}
\begin{equation}\label{met_gen}
  d s^2 \, = \, \frac{1}{z^2} \, \left[ - e^{a (\tau, z)} \, d \tau^2
    + \tau^2 \, e^{b (\tau, z)} \, d \eta_{st}^2 + e^{c (\tau, z)} \, d
    x_\perp^2 + d z^2\right]
\end{equation}
with $a (\tau, z)$, $b (\tau, z)$ and $c (\tau, z)$ some arbitrary
functions. One can show that in the $\eta_{st}$- and
$x_\perp$-independent case the most general EMT can be written as
\begin{equation}\label{emt_ads}
  \langle T^{\mu\nu} \rangle \, = \,
  \left( \begin{array}{cccc} \epsilon (\tau) & 0 & 0 & 0 \\
      0 & p (\tau) & 0 & 0 \\
      0 & 0 & p (\tau) & 0  \\
      0 & 0 & 0 & p_3 (\tau) \end{array} \right)
\end{equation}
at $\eta_{st} =0$, with the energy density $\epsilon (\tau)$, along with
the transverse and longitudinal pressures $p (\tau)$ and $p_3 (\tau)$.

Let us study the late-time asymptotics of this strongly-coupled
medium. Following \cite{Janik:2005zt} one assumes that at late times
the energy density of the medium scales as a power of proper time,
$\epsilon \sim \tau^\Delta$ with $-4 < \Delta <0$. One can then show
that in the $v \equiv z \, \tau^{\Delta /4} =$~const, $\tau
\rightarrow \infty$ limit the functions $a, b$ and $c$ depend on the
scaling variable $v$ only \cite{Janik:2005zt}. Solving the resulting
Einstein equations and requiring that there are no singularities in
the bulk, Janik and Peschanski \cite{Janik:2005zt} found that $\Delta
= -4/3$ and arrived at the following metric (cf. \peq{AdSSBH_FG})
\begin{equation}\label{JP_metric}
  ds^2 = \frac{L^2}{{z}^2}\left[ - \frac{\left( 1 - \frac{e_0}{3} \, 
 \frac{{z}^4}{\tau^{4/3}} \right)^2}{1 + \frac{e_0}{3} \,  \frac{{z}^4}{\tau^{4/3}}} \,
    d \tau^2 + \left( 1 + \frac{e_0}{3} \, \frac{{z}^4}{\tau^{4/3}} \right) \, 
\left( \tau^2 \, d \eta_{st}^2 + d x_\perp^{2}\right) + d {z}^2  \right],
\end{equation}
which is the gravity dual of the celebrated Bjorken hydrodynamics
\cite{Bjorken:1982qr}. Note that now $\epsilon (\tau) = 3 \, p (\tau)
= 3 \, p_3 (\tau) = (N_c^2 / 2 \pi^2) \, e_0/\tau^{4/3}$, in agreement
with \cite{Bjorken:1982qr} ($e_0$ is an arbitrary dimensionful
constant).  Since $\epsilon \sim T^4$ in this conformal plasma, we
conclude that $T \sim \tau^{-1/3}$, again in agreement with
\cite{Bjorken:1982qr}: the temperature falls off with time as the
system cools.  The metric \peq{JP_metric} has a horizon at
\begin{equation}
  \label{JPhor}
  z = z_h^{JP} \equiv \left( \frac{3}{e_0} \right)^{1/4} \ \tau^{1/3}.
\end{equation}
Comparing with \fig{AdSSBH_fig} we see that the horizon of the black
hole \peq{JP_metric} falls deeper into the bulk as the time goes on.

The important conclusion we draw from the result of
\cite{Janik:2005zt} is that a strongly-coupled $\eta_{st}$- and
$x_\perp$-independent ${\cal N} =4$ SYM medium would invariably end up
in a Bjorken hydrodynamics state. Hence one can derive the late-time
asymptotics of this medium without an explicit knowledge of the
medium's origin.

It is also interesting to study the approach to the Bjorken
hydrodynamics/Janik-Peschanski metric \peq{JP_metric}. This was done
in \cite{Nakamura:2006ih,Janik:2006ft} by expanding the coefficients
of the metric \peq{met_gen} around the scaling solution
\peq{JP_metric} for late times,
\begin{equation}\label{aaa}
  a (\tau, z) \, = \, a(v) + a_1 (v) \, \frac{1}{\tau^{2/3}} + a_2 (v)
  \, \frac{1}{\tau^{4/3}} + \ldots
\end{equation}
with similar expansions for $b (\tau, z)$ and $c (\tau, z)$. The
coefficients $a_1, a_2, b_1, \ldots$ were found by solving Einstein
equations perturbatively in $1/\tau^{2/3}$ with the matching near the
AdS boundary onto viscous hydrodynamics for the EMT of the gauge
theory. The latter yields the following EMT \cite{Danielewicz:1984ww}
\begin{equation}\label{emt_visc}
  \langle T^{\mu\nu} \rangle \, = \,
  \left( \begin{array}{cccc} \epsilon (\tau) & 0 & 0 & 0 \\
      0 & p (\tau) + \frac{2}{3} \, \frac{\eta}{\tau} & 0 & 0 \\
      0 & 0 & p (\tau)  + \frac{2}{3} \, \frac{\eta}{\tau} & 0  \\
      0 & 0 & 0 & p (\tau) - \frac{4}{3} \, \frac{\eta}{\tau} \end{array} \right)
\end{equation}
with $\eta$ now denoting shear viscosity. In a conformal medium where
$\epsilon \sim T^4$ the shear viscosity scales as $\eta \sim T^3 \sim
1/\tau$ in the Bjorken expansion. We hence write $\eta = (N_c^2 / 2
\pi^2) \, e_0^{3/4} \, \eta_0/\tau$ with $\eta_0$ a dimensionless
constant. Requiring that the resulting metric has no singularities in
the bulk yields \cite{Janik:2006ft}
\begin{equation}
  \label{eta0}
  \eta_0^2 \, = \, \frac{\sqrt{3}}{18}
\end{equation}
such that the shear viscosity is \cite{Janik:2006ft,Heller:2007qt}
\begin{equation}
  \label{eta1}
  \eta = \frac{N_c^2}{2^{3/2} \, 3^{3/4} \, \pi^2} \, \frac{e_0^{3/4}}{\tau}. 
\end{equation}
Comparing the EMT of a SYM plasma \peq{EMT_plasma} with $\epsilon =
(N_c^2 / 2 \pi^2) \, \epsilon_0/\tau^{4/3}$ we read off the
temperature
\begin{equation}
  \label{Temp}
  T (\tau) = \frac{2^{1/2}}{3^{1/4}} \, \frac{e_0^{1/4}}{\pi \, \tau^{1/3}}
\end{equation}
which, when used in \eq{eta1} gives
\begin{equation}
  \label{shear_v}
  \eta = \frac{\pi}{8} \, N_c^2 \, T^3. 
\end{equation}
This important relation between the shear viscosity and temperature of
the strongly-coupled ${\cal N} =4$ SYM plasma was derived originally
by Kovtun, Policastro, Son, and Starinets
\cite{Policastro:2001yc,Son:2002sd,Policastro:2002se,Kovtun:2003wp,Kovtun:2004de}
using the Kubo formula to relate $\eta$ to the absorption cross
section of a graviton by the AdSSBH.

Note that combining Eqs. \peq{shear_v} and \peq{entr_strong} one gets
\cite{Policastro:2001yc,Son:2002sd,Policastro:2002se,Kovtun:2003wp,Kovtun:2004de}
\begin{equation}
  \label{eta/s}
  \frac{\eta}{s} = \frac{1}{4 \, \pi},
\end{equation}
where $s = S/V_{3d}$ is the entropy density. Based on the consistency
of this result for a variety of dual geometries it has been
conjectured in \cite{Kovtun:2003wp,Kovtun:2004de} that the ratio
\peq{eta/s} is a universal lower bound on $\eta/s$ for any theory
(including QCD):
\begin{equation}
  \label{bound}
  \frac{\eta}{s} \ge \frac{1}{4 \, \pi}.
\end{equation}
This is known as the Kovtun-Son-Starinets (KSS) bound. It was proven
for theories with gravity duals in \cite{Kovtun:2004de}.  The
possibility of a violation of this bound was discussed in
\cite{Brigante:2007nu,Kats:2007mq,Buchel:2011uj}. The question of the
value of $\eta/s$ in QCD has been extensively studied in lattice QCD
\cite{Meyer:2007ic} and by viscous hydrodynamics simulations for heavy
ion collisions
\cite{Teaney:2003kp,Romatschke:2007mq,Song:2010mg,Qiu:2011hf}. The
current estimates place $\eta/s$ in QCD (extracted from heavy ion
collisions data) very close to the KSS bound \cite{Qiu:2011hf}, though
one has to remember that QCD is not a conformal theory, and $\eta/s$
is $T$-dependent in QCD unlike ${\cal N} =4$ SYM \cite{Meyer:2007ic}.

Returning to the time evolution of the strongly-coupled
rapidity-independent medium, note that further higher-order
$1/\tau^{2/3}$ corrections have been calculated in
\cite{Heller:2007qt,Benincasa:2007tp}. Reversing the problem, one can
ask about the possible early-time asymptotics of this medium. It was
shown in \cite{Kovchegov:2007pq} that for $\epsilon \sim \tau^\Delta$
only even integer $\Delta$ are allowed as $\tau \to 0$. The
non-negativity of the energy density demands that $\Delta =0$
\cite{Kovchegov:2007pq,Beuf:2009cx}. In such a scenario the energy
density $\epsilon$ starts out as a constant in time and eventually
begins to fall off like $\epsilon \sim \tau^{-4/3}$ as dictated by
Bjorken hydrodynamics. [Note however that if one expands Eqs.
\peq{T++}, \peq{T--}, and \peq{T+-} to the lowest order in $\tau$ (as
$x^\pm = \tau e^{\pm \eta_{st}}/\sqrt{2}$) one would obtain $\epsilon
\sim \tau^2$ \cite{Grumiller:2008va}, such that $\Delta =2$, though
the medium produced in a shock wave collision is not boost invariant.]
For numerical simulations of rapidity-independent (boost-invariant)
plasmas see \cite{Heller:2011ju,Chesler:2009cy}.

The weakness of the approach presented in this Section is at the early
times: one has to specify the metric to be used as the initial
condition for Einstein equations at some small value of $\tau =
\tau_0$. The EMT of the particles formed in the early stages of a
heavy ion collision would only specify $g_{\mu\nu}^{(4)}$ at $\tau_0$
in the expansion of the metric near $z =0$. Higher order metric
coefficients $g_{\mu\nu}^{(i)}$, $i \ge 6$, have to be specified too
at $\tau_0$: in the AdS/CFT dictionary they are likely to be dual to
some higher-dimensional operators in ${\cal N} =4$ SYM theory. It
appears that to insure the matching between the (presumably)
weakly-coupled early-time dynamics and the later-time dynamics
described by AdS/CFT one has to specify an infinite tower of
expectation values of operators at $\tau_0$ resulting from the former
to be used to initiate the latter.  To date the problem of how to do
this in a both rigorous and realistic way remains open.


\section{Jet quenching in a strongly-coupled medium}

Many experimental signals of QGP formation and evolution can be
calculated in the framework of AdS/CFT duality. Above we have
discussed the implications of the AdS/CFT correspondence for the
hydrodynamic description of heavy ion collisions. In the spirit of a
brief review, we will only mention one more signal of QGP: jet
quenching. The idea that jets can lose energy in QGP which would lead
to their suppression was originally proposed by Bjorken in
\cite{Bjorken:1982tu}, with the first calculations of jet quenching in
the perturbative QCD framework carried out in
\cite{Gyulassy:1993hr,Baier:1996sk,Zakharov:1996fv,Gyulassy:2000er}.

In the AdS/CFT framework the question of jet energy loss was first
tackled in \cite{Liu:2006ug} with the goal of determining a
strong-coupling value of the jet quenching parameter $\hat q$
originally defined in \cite{Baier:1996sk} in the perturbative QCD
framework. The purely AdS approach to jet quenching was first
advocated in
\cite{Gubser:2006bz,CasalderreySolana:2007qw,Herzog:2006gh,Sin:2004yx,Chernicoff:2006hi,Horowitz:2009pw}
and involved calculating the drag force on a heavy quark moving
through the SYM plasma. At zero temperature a heavy quark is dual in
AdS to the endpoint of an open Nambu-Goto string terminating on a D7
(flavor) brane wrapped around S$^5$, with the other end of the string
stretching to the stack of D3 branes at $z=\infty$
\cite{Karch:2002sh}. For a finite temperature medium, if we work in
the coordinate system employed in writing down the metric
\peq{AdSSBH}, we can only follow the string down to the AdSSBH
horizon.

\begin{figure}[h]
  \begin{center}
    \includegraphics[width=9cm]{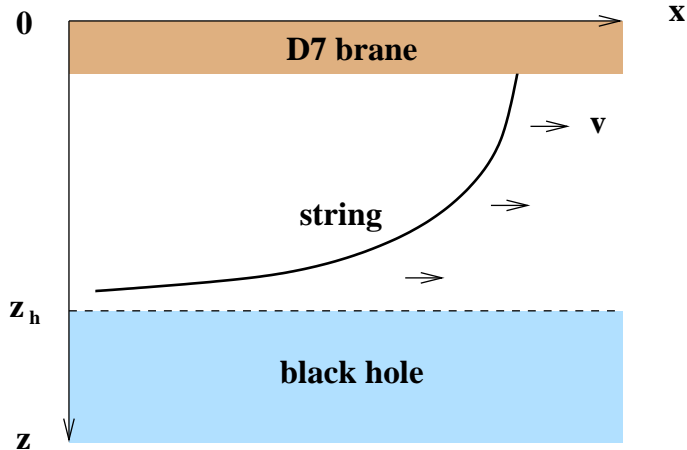}
  \end{center}
  \caption{A sketch of the trailing string configuration dual to a heavy 
    quark propagating in the strongly-coupled ${\cal N} =4$ SYM
    plasma.}
  \label{quench}
\end{figure}

Let us find the drag force on a heavy quark moving with a constant
speed $v$ through a supersymmetric plasma (assume that some external
force is applied to keep the quark velocity constant). To do this, one
first has to find the string configuration in the AdSSBH background
\peq{AdSSBH} dual to the moving quark. This is done by extremizing the
Nambu-Goto action of an open sting
\begin{equation}\label{NG}
  S_{NG} = - \frac{\sqrt{\lambda}}{2 \, \pi \, L^2} \, 
\int \, d\tau \, d\sigma \, \sqrt{- \mbox{det} \, g_{ab}}, \ \ \ \ \
  g_{ab} = g_{MN} \,  \partial_a X^M \, \partial_b X^N
\end{equation}
with the boundary condition requiring that the string endpoint
attached to the D7 brane moves with velocity $v$ along the $x$-axis.
Here $g_{MN}$ is the AdSSBH metric \peq{AdSSBH} and $X^M=X^M (\tau,
\sigma)$ are sting coordinates: they specify the mapping from the
string world-sheet coordinates $\sigma^a = (\tau, \sigma)$ with $a= 0,
1$ to spacetime coordinates $x^M$. (Note that $\partial_a =
\partial/\partial \sigma^a$.) The resulting classical string
configuration is
\begin{equation}
  \label{config}
  X^M (\tau = t, \sigma =z) \, = \, \left( t, 0, 0, x(t,z), z \right)
\end{equation}
with \cite{Gubser:2006bz,CasalderreySolana:2007qw,Herzog:2006gh}
\begin{equation}
  \label{string}
  x (t, z) = v \, t  - \frac{z_h \, v}{2} \, \left[ \arctan \frac{z_h}{z} 
  + \frac{1}{2} \, \ln \left( \frac{z_h + z}{z_h -z} \right) - \frac{\pi}{2} \right]. 
\end{equation}
It is illustrated in \fig{quench}. The drag force (momentum change) is
given by the momentum flow down the string from its endpoint attached
to the D7 brane,
\begin{equation}
  \label{mom-flow}
  \frac{dp}{dt} = - \pi^1_x
\end{equation}
where $p$ is the 3-momentum of the quark and $\pi^1_x$ is the
canonical momentum of the string defined by
\begin{equation}
  \label{canon}
  \pi^a_\mu = - \frac{\sqrt{\lambda}}{2 \, \pi \, L^2} \, 
\frac{\partial\sqrt{-\mbox{det} \, g_{bc}}}{\partial (\partial_a X^\mu)}.
\end{equation}
Using Eqs. \peq{string}, \peq{config}, \peq{canon} in \eq{mom-flow}
gives us the drag force on a heavy quark in the strongly coupled
${\cal N} =4$ SYM plasma
\cite{Gubser:2006bz,CasalderreySolana:2007qw,Herzog:2006gh}
\begin{equation}
  \label{drag}
  \frac{dp}{dt} = - \frac{\pi \, \sqrt{\lambda}}{2} \, T^2 \, \frac{p}{M}
\end{equation}
where $M$ is the mass of the heavy quark. (The mass appears since $p/M
= v/\sqrt{1-v^2}$.) \eq{drag} was derived for a static medium: it was
generalized to the case of Bjorken hydrodynamics corresponding to the
Janik-Peschanski metric \peq{JP_metric} in \cite{Giecold:2009wi} with
the conclusion that \eq{drag} still applies in this dynamic case if
one replaces $T \to T(\tau)$ in it, with $T(\tau)$ given by \eq{Temp}.

The AdS/CFT result for the drag force \peq{drag} allowed for some
interesting phenomenology of the heavy quark energy loss at RHIC
\cite{Horowitz:2007su}. However, one has to be careful applying this
large-coupling result to QCD jet physics: due to asymptotic freedom,
at least some part of the jet coupling to the medium should be
perturbative. On top of that, the model considered here involves a
quark which is being dragged through the medium by an external force
which prevents it from slowing down, while in real life the hard
partons simply plow through the medium loosing energy and momentum.
Modeling of jets in the medium along these more realistic lines was
carried out in \cite{Hatta:2008tx}. The conclusion was that, due to
strong coupling, the partons are highly likely to branch into more
partons, distributing the energy democratically between them. In the
end one obtains an isotropic distribution of partons, very different
from the jet cone of perturbative QCD and from the jet cones seen in
the actual collider experiments (see also \cite{Hofman:2008ar}).  This
difference between the ``AdS jets'' and real life jet gives us another
argument in favor of being cautious in applying AdS results for jet
quenching to the real-life heavy ion collisions.

Applications of the AdS/CFT correspondence to jet quenching are by no
means limited by the calculation we have just presented. Further
research addressed issues of studying Mach cones generated by the
supersonic jets, energy loss of light quarks and gluons, momentum
broadening of hard partons traversing the medium, quark and gluon
stopping distance, dissociation and melting of mesons, along with
other topics.  A proper discussion of these important results lies
outside of the scope of this brief review. We refer the interested
reader to \cite{CasalderreySolana:2011us} for a detailed review of
these subjects.


\section{Conclusions}

Our brief review covers several topics on applications of AdS/CFT
correspondence to heavy ion collisions, with an emphasis on modeling
the actual collisions, stemming in part from the author's subjective
interests and research experience. Indeed, due to the compact format
of this review, many important topics in applications of AdS/CFT
correspondence to heavy ion collisions had to be omitted. Among those
topics are two-particle correlations (see \cite{Son:2002sd} for a
pedagogical presentation of their calculation), electromagnetic probes
\cite{CaronHuot:2006te}, heavy quark potential at finite temperature
\cite{Rey:1998bq,Brandhuber:1998bs,Bak:2007fk,Albacete:2008dz,Grigoryan:2011cn},
jet quenching-related observables listed above, and efforts to
calculate the elliptic flow $v_2$ in AdS \cite{Taliotis:2010pi}.

We hope that we have convinced the reader that practically any heavy
ion observable can be modeled using the AdS/CFT correspondence,
generating interesting qualitative and, sometimes, quantitative
insight. In comparing the results of AdS/CFT calculations to the
actual heavy ion data or to lattice QCD simulations one has to be
careful to remember the many differences between QCD and ${\cal N} =
4$ SYM theory, though one hopes that certain universal features are
common for both theories. One may also hope that future research would
put these differences between QCD and ${\cal N} = 4$ SYM under
quantitative theoretical control.


\section*{Acknowledgments}

I am grateful to Ulrich Heinz, Hovhannes Grigoryan, and Samir Mathur
for the discussions which greatly helped me in writing this review.

This work is sponsored in part by the U.S. Department of Energy under
Grant No. DE-SC0004286.


\section*{References}


\providecommand{\href}[2]{#2}\begingroup\raggedright\endgroup

\end{document}